\documentclass[aip,
 cp, 
 amsmath,amssymb,
 reprint,%
]{revtex4-2}

\usepackage{hyperref} 
\usepackage{graphicx}
\usepackage{dcolumn}
\usepackage{bm}
\usepackage{upgreek}
\usepackage{textcomp}
\usepackage[utf8]{inputenc}
\usepackage[T1]{fontenc}
\usepackage{bbm}

\usepackage{mathptmx} 

\begin{document}

 
\title{Practical Doppler broadening thermometry}

\author{Nicola Agnew} 
 \email[Corresponding author: ]{nicola.agnew@strath.ac.uk}
 \affiliation{
  Department of Physics, SUPA, University of Strathclyde, 107 Rottenrow E, Glasgow, G4 0NG, Scotland, UK 
.
}
 \author{Graham Machin}
\affiliation{%
National Physical Laboratory, Hampton Rd, Teddington, TW11 0LW,  UK 
}%
\author{Erling Riis}%
 \author{Aidan S. Arnold}%
 \email{aidan.arnold@strath.ac.uk}
\affiliation{
  Department of Physics, SUPA, University of Strathclyde, 107 Rottenrow E, Glasgow, G4 0NG, Scotland, UK 
.
}

\date{\today} 

\begin{abstract}
 We report initial research to develop a compact and practical primary thermometer based on Doppler broadening thermometry (DBT). The DBT sensor uses an intrinsic property of thermalized atoms, namely, the Doppler width of a spectral line characteristic of the atoms being probed. The DBT sensor, being founded on a primary thermometry approach, requires no calibration or reference, and so in principle could achieve reliable long-term \emph{in-situ} thermodynamic temperature measurement. Here we describe our approach and report on initial proof-of-concept investigations with alkali metal vapour cells. Our focus is to develop long-term stable 
 thermometers based on DBT that can be used to reliably measure temperatures for long periods and in environments where sensor retrieval for re-calibration is impractical such as in nuclear waste storage facilities.
\end{abstract}

\maketitle

\section{\label{sec:intro} Introduction}

The \emph{mise en pratique} for the definition of the kelvin (MeP-K-19 \cite{Fellmuth2016}) has opened new possibilities for practical primary thermometry as a means of providing in-situ traceability for temperature -- 
a long-term objective for the international thermometry community. Currently the overwhelming majority of thermometry approaches (resistance-based thermometers and thermocouples) require periodic traceable calibration against the International Temperature Scale of 1990 (ITS-90) \cite{PrestonThomas1990} to maintain reliable operation -- which is often costly and time consuming. As a result there has been an increasing international effort in developing practical primary thermometry standards to be utilised in industry. One approach to providing in-situ temperature traceability, Johnson Noise Thermometry, is already well advanced \cite{Bramley2020}, whilst a number of photonic based approaches are under active investigation \cite{Dedyulin2022}.

Doppler broadening thermometry (DBT) 
is a primary photonic-based temperature measurement scheme that capitalizes on the Doppler effect. While it's not yet officially recognized in the MeP-K-19, DBT's initial demonstration for determining the Boltzmann constant in 2005 using NH$_3$ molecules marked a significant achievement in the field of temperature measurement\cite{2005lasp.conf..104D}. Many of the subsequent DBT demonstrations \cite{PhysRevLett.111.060803,Gianfrani2012,Gianfrani2016,Gotti2018} have largely focused on the study of molecular transitions. Although results have been promising, with uncertainties as low as 24 parts per million achieved \cite{PhysRevLett.111.060803,Gotti2018},  this approach encounters significant collisional perturbations and the low optical cross-sections of the molecules require large path lengths to obtain a sufficient signal-to-noise ratio. This can hinder scalability and limit the practical applications of Doppler broadening thermometry.

An alternative approach to DBT, using 
alkali metal vapours \cite{Truong2011,Truong2015,Pan:19}, offers several advantages over molecular species. Alkali atoms possess a relatively simple atomic structure, such that more complex effects, such as rotational state-changing collisions, do not need to be accounted for. Additionally, these atoms exhibit a high optical cross-section, enabling operation at low pressures with minimal collisional broadening. This attribute allows for the use of smaller vapour cells, which is desirable for practical implementation and scalability of DBT techniques. However, 
DBT using atoms presents its own set of challenges \cite{Truong2011}, for example the large optical cross section leads to optical pumping effects so care must be take to perform measurements within the weak probe regime \cite{Sherlock2009}. Additionally, the atomic structure is highly susceptible to external magnetic fields, requiring the use of magnetic shielding to minimize their impact on measurements.

In this paper, we present our approach and preliminary findings for using alkali metal vapor cells for DBT. Specifically, we conducted spectroscopic investigations on the Rubidium (Rb) D$_2$ line at room temperature and examined the impact of amplified spontaneous emission (ASE) from a laser diode on the measured absorption spectrum. This initial work is aimed at addressing the challenge of accurately measuring absorption spectra in the presence of ASE, a common issue in spectroscopy, via a combination of experimental techniques and post-measurement processing.

With our proposed methodology, we successfully mitigate, through the use of appropriate corrections, the adverse effects of ASE on our spectroscopic data. It should be noted that this work primarily focuses on the absorption of the spectra rather than the determination of the Doppler width itself. This idea of absorption thermometry, which underpins our current methodology, relies on deriving temperature from the measured absorption using the Beer-Lambert law. Although this method of temperature derivation is dependent on the vapour cell length and thus not inherently primary in nature, we posit that this technique can serve as a valuable check on our data when we begin extracting Doppler broadening thermometry derived temperatures. 

Furthermore, we build upon our findings by theoretically investigating the Rb D$_2$ line and evaluate its potential suitability for temperature measurement. To achieve this we employ the computational theoretical program  \textit{ElecSus} \cite{Zentile2015,Keaveney2018,Siddons2008} which models the absorption spectrum of ensembles of atoms. We study the variation of the optical absorption as a function of temperature and cell length. Our analysis offers insight into finding the most favourable potential locking point for obtaining the optimal temperature reading using the characteristic absorption of the vapour at a given frequency. Our study shows the effects of temperature and cell length on the absorption spectrum, providing beneficial information for optimizing the design and operation of alkali vapor cell-based temperature sensors which give a temperature readout via a combination of measurement of the absorption co-efficient and the Doppler width. 

\maketitle

\section{\label{sec:theory} Theory}

It is important to model the underlying physics governing the interaction between the Rb vapour and the laser system, to obtain a basis for comparison for our experimentally obtained results. This model takes a given temperature and vapour cell length as input parameters and generates the predicted spectroscopy signal as an output. 
Here the Rb vapour cell is modelled as a dielectric material with a refractive index 
$n$, which is probed by a laser with an angular frequency 
$\omega$. The atoms' optical response is determined by the electric susceptibility $\chi$, which is a function of the laser's detuning $\delta_j$, from the atomic resonance $\omega_j$ of the $j^\textrm{th}$ transition, $(\delta_j=\omega-\omega_j)$. The electric susceptibility comprises both real and imaginary parts, where the real part characterizes the atoms' dispersion profile, and the imaginary part characterizes the absorption profile - related via the Kramers-Kronig relations \cite{deLKronig1926}. The imaginary part of the susceptibility of the $j^\textrm{th}$ transition is calculated such that:
\begin{equation} \label{eq:1}
\chi(\delta_j)=\frac{C_j^{2}d^{2}N_a}{\epsilon_0\hbar}f(\delta_j),
\end{equation}
where $\epsilon_0$ denotes the permittivity of free space, $\hbar$ is the reduced Planck's constant, $d$ the dipole matrix element, $f(\delta_j)$ the lineshape function, and $C_j$ the relative strength factor, a quantity which is dependent on the excited and ground state of the relative hyperfine transition - see \cite{Zentile2015,Siddons2008} for further information on the calculation of this term. The atomic number density is denoted as  $N_{\textrm{a}}$ and is calculated as follows:
\begin{equation} \label{eq:2}
N_{\textrm{a}} = \frac{F_{\textrm{a}}\,N}{2(2I+1)},
\end{equation}
where $F_{\textrm{a}}$ is the isotopic abundance (\(F_{\textrm{a}}\approx\) $72\,$\% for $^{85}$Rb and \(F_{\textrm{a}}\approx\) $28\,$\% for $^{87}$Rb) \cite{85RbSteck,87RbSteck}, $I$ denotes the nuclear spin quantum number ($I=5/2$ for $^{85}$Rb and $I=3/2$ for $^{87}$Rb) and $N$ the elemental number density. 

The relation in Eq.~(\ref{eq:2}) is valid under the assumption that the temperature of the atoms is low and the probe beam is weak such that the atoms are evenly distributed among the ground state manifold. The elemental number density can be calculated as $P/(kT)$, wherein $P$ denotes the pressure, $T$ the temperature and $k$ the Boltzmann constant. The pressure is derived from the vapour-pressure model for Rb \cite{85RbSteck,87RbSteck} which is known to give an accuracy of better than $5\,$\% \cite{85RbSteck}. With adequate temperature calibration, our results may lead to an improved vapor-pressure model. 

The lineshape $f(\delta_j)$ of the imaginary susceptibility shown in Eq.~(\ref{eq:1}) is Lorentzian. The atoms experience a spontaneous decay at a rate characterized by \(\Gamma\) from excited state to ground state. This leads to a natural broadening of the transition, given by the complex function:
\begin{equation} \label{eq:4}
    f(\delta_j)=\frac{\mathbbm{i}}{\Gamma/2-\mathbbm{i}\delta_j}.
\end{equation}

In addition to the homogeneous natural broadening in Eq.~(\ref{eq:4}), the thermal motion of atoms in the vapor induces a frequency shift of the interacting laser light proportional to their velocity $v$, leading to inhomogeneous broadening. This phenomenon, known as Doppler broadening, gives rise to a Gaussian lineshape in the frequency domain. The absorption profile of atoms in the Doppler regime is commonly modeled using a Voigt profile, which is a convolution of a Gaussian and a Lorentzian lineshape. This can be calculated rigorously by integrating over a Maxwell-Boltzman distribution convoluted with the Lorentzian $f(\delta_j-k_\textrm{L} v)$, where $k_\textrm{L}$ is the wavevector of the laser radiation. In doing so, we obtain: 
\begin{equation} \label{eq:5}
    V(\delta_j) = \frac{C_j^{2}d^{2}N_a}{\epsilon_0\hbar\sqrt{\pi} u}\int_{-\infty}^{\infty}f(\delta_j-kv)e^{-v^2/u^2}dv,
\end{equation}
where $u=\sqrt{2kT/m}$ is the mean speed    
for an atom of mass $m$ at temperature $T$.
Rather than employing integration, the Voigt profile can be approximated using the complex error function \cite{Schreier1992}. This method decreases the computational time required for modeling the absorption spectrum, but at the cost of less reliable lineshape fitting. 

\maketitle
\section{\label{sec:expt} Data Acquisition and processing}

\begin{figure}[!b]
\includegraphics[width=1\columnwidth]{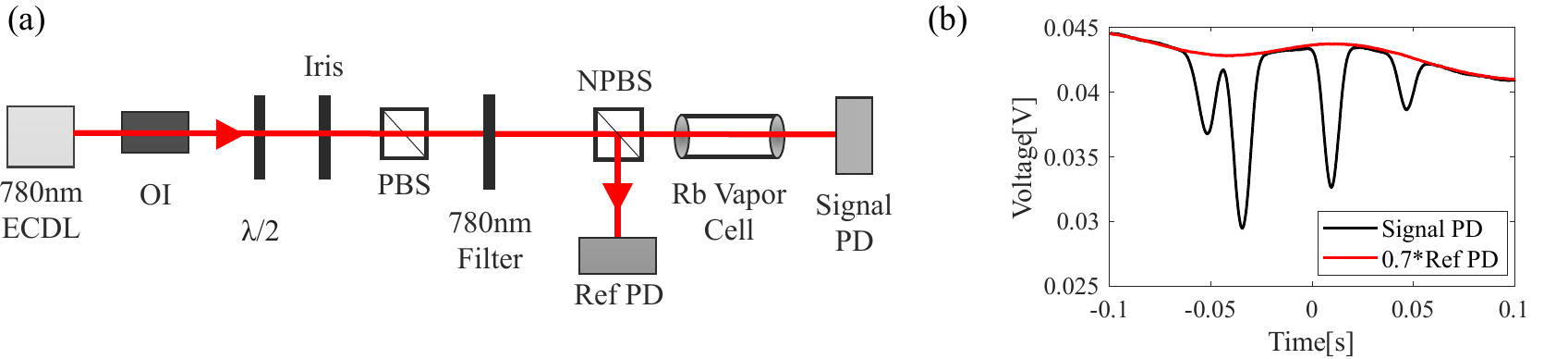}
\caption{\label{Fig:setup} (a) Experimental setup for DBT. ECDL (external cavity diode laser), OI (optical isolator), PBS (polarizing beam splitter), NPBS (non-polarizing beam splitter) PD (photodiode). (b) Raw data collected from the reference photodiode (red) and signal PD (black). }
\end{figure}

The experimental setup utilized to acquire spectroscopy data is illustrated in Fig.~\ref{Fig:setup} (a), with the resulting raw data shown in Fig.~\ref{Fig:setup} (b). The setup consists of an external cavity diode laser (ECDL) \cite{Arnold1998,Daffurn2021} that is frequency scanned across the Rb D$_2$ line at $780\,$nm. The laser output is directed through an iris of diameter $1.8\,$mm to reduce the ASE. A halfwave plate and polarizing beam splitter are employed  
to control the laser output power. The beam is then divided using a non-polarizing beam splitter, with one output serving as the reference signal and the other sent into a $74\,$mm Rb vapour cell, where it is detected using a silicon-based photodiode (PD). The vapour cell laser power is set below $1\,\upmu$W to minimise power broadening and optical pumping effects, which have not been considered in the theoretical model \cite{Zentile2015,Keaveney2018}. In this experiment the Rb cell temperature is assumed to be that of the room, whose temperature is recorded to be $297.2\,$K using a CURCONSA CO$_2$ meter with a built-in temperature sensor. In future work direct measurement of the cell's temperature will be performed, and in addition temperature control of the cell will be implemented to test the limits of the spectroscopy data for deriving temperature in both warmer and colder environments.

The spectral data is acquired and then processed in MATLAB. Initially, the mean of ten successive points in the spectral data is calculated to minimize the presence of noise in the data. Following this, the power of the laser is computed by determining the mean voltage of the reference signals and scaling it by a factor of 0.7 to align it with the spectroscopy PD signal - as illustrated in Fig.~\ref{Fig:setup} (b). The average current flowing through the PD was subsequently determined using the PD's resistance, and the power was calculated using the wavelength-dependent responsivity value specified in the PD data sheet. 
Subsequently, the transmission axis of the data was defined in the range of 0 to 1 through normalization. However, as demonstrated in Fig.~\ref{Fig:setup} (b), the maximum transmission is non-linear \cite{Pizzey2022}. Although this non-linearity can be partially corrected by dividing the spectroscopy signal by the reference signal before normalization, the overlap of these signals is not perfect, and some non-linearity persists. A secondary correction method for power fluctuations in the observed data involves fitting a polynomial to points on 
the reference scan transmission and normalising.  
This works best for data with a long scan range.

To initially calibrate the time of the scan into frequency units, the time difference of two of the maximum absorption points in the data is quantified, then scaled with its theoretical frequency counterpart. To aid this, we use the location of the minimum of a parabolic fit to the bottom $2\,$\% of the absorption peaks -- to avoid data noise giving false minima.

Upon completion of normalization, it is necessary to eliminate the ASE from the data, by subtracting the transmission values at the floor of warm cell data (see Fig.~\ref{Fig:ASE}) from those of the room temperature data. 
The transmission of the Rb cell was therefore measured after heating with a heat gun to a high optical density. The heating was carried out until the spectroscopy absorption signal became saturated, as evidenced by a flat transmission floor, as shown in Fig.~\ref{Fig:ASE}. 
The objective was to quantify the level to which the 
ASE from the laser diode, as well as stray light and current noise from the PD, contributed to an offset in transmission signal \cite{Daffurn2021,Pizzey2022}. Without these elements, in theory, the signal floor would ideally be at $0\,$\% transmission, indicating that all resonant laser light is absorbed by the atoms. The ASE produced by the laser, which is off resonance, passes through the cell to the PD, leading to an apparent increase in the transmission floor. The ASE is quantified by taking the mean value of a linear fit across the signal floor. Quantifying the contribution of ASE in the data is critical as it affects the signal's amplitude and should be suppressed as much as possible. To reduce ASE, we incorporated a wavelength-dependent band-pass filter with a high degree of out-of-band blocking into the setup.

Discrepancies in the frequency domain arise between the experimental data and theory because the frequency sweep exhibits non-linear behaviour whereas the theoretical framework assumes a linear scanning process. To address this, we apply a cubic polynomial mapping technique to correct the theory. Specifically, we utilize parabolic fits to the minima in transmission of the four Doppler absorption peaks observed in the experimental data as known values, and determine the unknown parameters by solving a system of four simultaneous equations expressed in the form:
\begin{equation} 
\label{eq:6} 
\delta_\textrm{\,t}=A+B\delta_\textrm{e}+C{\delta_\textrm{e}}^{2}+D{\delta_\textrm{e}}^{3}, 
\end{equation}
where $A$, $B$, $C$ and $D$ are our unknown values and the subscript `e' refers to experimental data, and `t' denotes theoretical values.

\maketitle

\section{\label{sec:exptresults}Experimental Results}

\begin{figure}[!b]
\includegraphics[width=0.8\textwidth]{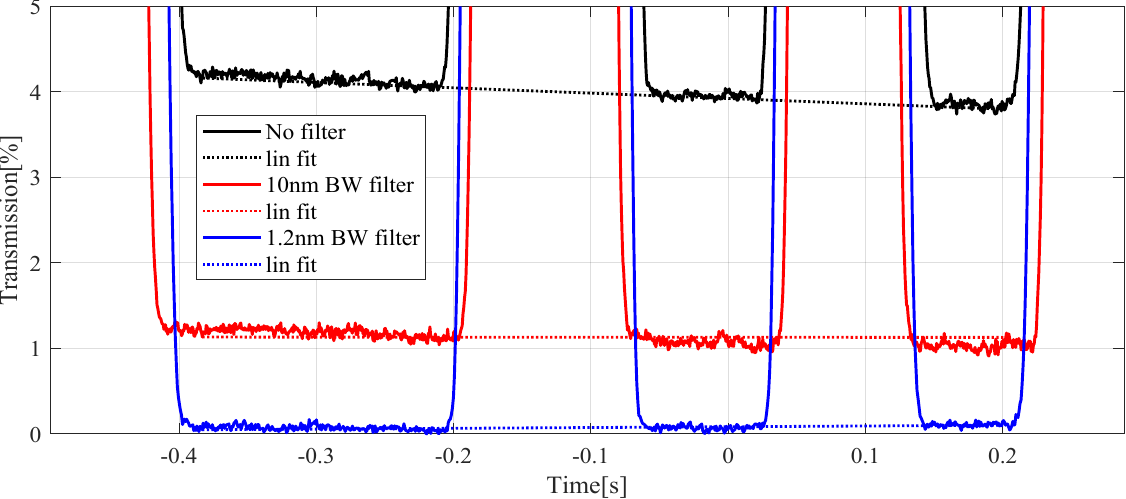}
\caption{\label{Fig:ASE} Warm cell measurements of the Rb D$_2$ line, the black curve shows the ASE for when no filter is used, the red curve shows the ASE levels when employing a $10\,$nm BW filter and the blue curve the ASE with a $1.2\,$nm BW filter.}
\end{figure}

The outcomes of the warm cell measurements of the Rb D$_2$ line are presented in Fig.~\ref{Fig:ASE}. The spectrum was captured under three different conditions: 
without any filter; with a $780$-$10\,$nm  filter; and using a $780$-$1.2\,$nm filter. We use the notation CW-BW (Centre wavelength - Bandwidth (BW) full-width-at-half-maximum) to describe CW-BW band-pass interference filters. Without a filter the average ASE level of the ECDL was observed to be $(4.0\pm0.3)\,$\%. With the $780$-$10\,$nm and $780$-$1.2\,$nm filters, the ASE level decreased to $(1.1\pm0.1)\,$\% and $(0.08\pm0.02)\,$\%, respectively -- up to a 50-fold improvement. Suppression of ASE is of utmost importance for accurate temperature measurement -- significant ASE levels can introduce a deceptive increase in the transmission of the spectra. Consequently, when employing ASE-sensitive techniques for temperature measurement -- such as absorption thermometry -- high ASE levels can yield erroneous readings that underestimate the actual temperature.  
Minimising ASE maintains the integrity and reliability of the temperature measurement and minimises the respective uncertainty budget. 

\begin{figure}[!t]
\includegraphics[width=0.9\textwidth]{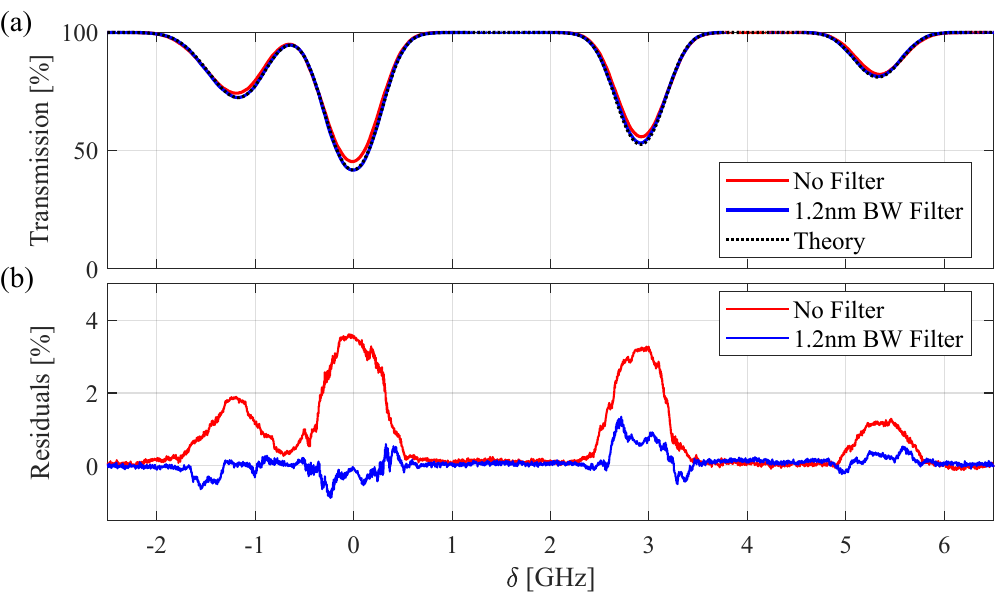}
\caption{\label{Fig:spectrum} (a) The absorption spectrum of the Rb D$_2$ line at room temperature of approximately $297.2\,$K for a cell length of $74\,$mm. The theoretically calculated spectrum for these values is shown in black. The spectrum measured experimentally using a $1.2\,$nm BW filter at $780\,$nm is shown in blue and without the filter in red, The residuals of both data sets are shown in (b)}
\end{figure}

The resulting experimentally obtained spectra and 
residuals for the Rb D$_2$ line at the measured room temperature ($297.2\,$K) are show in Fig.~\ref{Fig:spectrum}. We observe a significantly improved agreement between the experimental results and the theoretical prediction when a narrowband filter with a BW of $1.2\,$nm is employed to mitigate the effects of ASE. By applying the post processing techniques outlined previously to remove what remains of the ASE in the data obtained with the narrowband filter, a scan root-mean-square (RMS) error of $0.33\,$\% was calculated over the full scan range of -2.5 to $6.5\,$GHz detuning. For the data obtained without the filter, and $\approx4\,$\% ASE, we obtained an RMS error of $1.29\,$\%. This four-fold improvement in the agreement with theoretical predictions underscores the importance of employing a  narrowband filter to reduce the detrimental impact of ASE on the measurement.

The experimental data was fitted against the theoretical model to ascertain the temperature at which the RMS error is at its lowest. We do so by systematically varying the models temperature until the RMS error reaches a minimum value. Following this, we defined the temperature error as the difference between the theoretically determined temperature that yields the lowest RMS error and the temperature measured in the actual experimental setting. It is found that for the data taken without the filter the RMS error is the lowest ($0.212\,$\%) at a temperature $296.4\,$K, corresponding to a temperature error of $(0.8\pm0.1)\,$K. Experimental data taken with the $1.2\,$nm BW filter has the best agreement with theoretical data at $297.2\,$K, corresponding to a RMS error of $0.268\,$\% and a temperature error of $(0.0\pm0.1)\,$K. Overall these results show that the ASE from the laser system being implemented can significantly affect the accuracy of the observed spectral features. Furthermore, characterising the ASE and using the appropriate filtering techniques can minimise the detrimental impact of ASE. 

\begin{figure}[!t]
\includegraphics[width=1\textwidth]{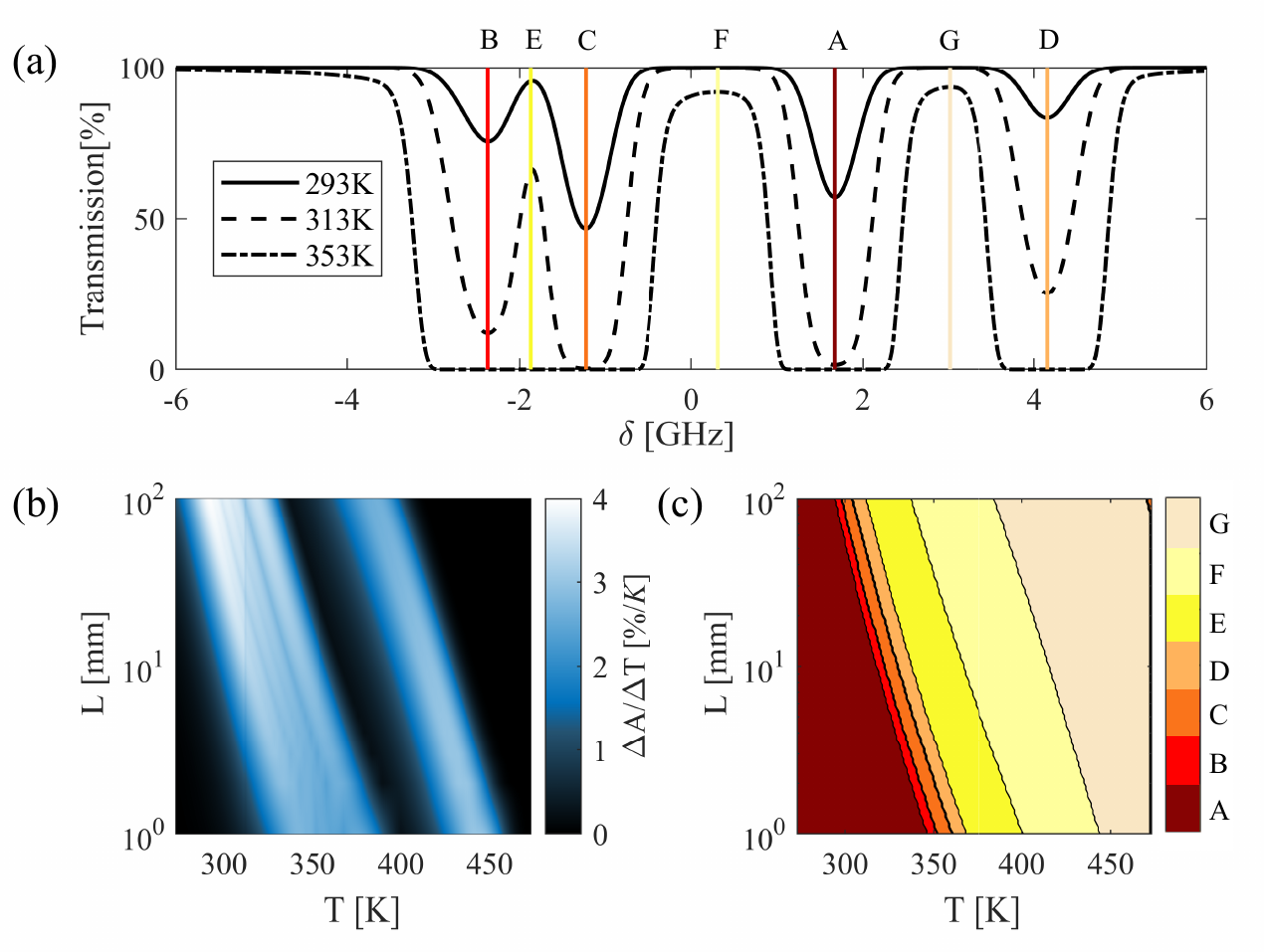}
\caption{\label{Fig:Rb} The theoretically calculated Rb D$_2$ line absorption spectrum is presented in (a) for three temperatures, with a fixed cell length of $74\,$mm. In (b), we illustrate the  percentage change in absorption per kelvin of temperature variation, \textit{maximised} over the seven specific detuning values (A-G) indicated by the vertical lines in (a). 
The detuning value (A-G marked in (a)) corresponding to the maximum transmission change with temperature at a given temperature and cell length (shown in (b)) is depicted in (c).
}
\end{figure}

\section{\label{sec:model}Modelling Results}

The absorption spectra of the Rb D$_2$ lines were theoretically modeled to investigate the percentage change in transmission per kelvin of temperature variation. The temperature range examined was from $273\,$K to $473\,$K, and the cell lengths ranged from $1\,$mm to $100\,$mm. The analysis focused on determining the percentage change at the transmission minima (labelled A, B, C and D in Fig.~\ref{Fig:Rb} (a)) --  as well as at the transmission maxima (labelled E, F and G). 

The transmission minima saturate at elevated temperatures, which can be attributed to the increase in optical depth due to increased vapour pressure. At these saturation points, determination of the temperature by transmission becomes unfeasible. Nevertheless, in principle, the laser can be locked to any desired frequency, and points of maximum transmission could be utilized as an alternative temperature indicator at higher temperature regimes. 

The percentage change in transmission per kelvin temperature change is shown in Fig.~\ref{Fig:Rb}(b) and the corresponding potential `locking' point demonstrating the highest sensitivity to temperature variation (out of those marked A-G on Fig.~\ref{Fig:Rb}(a)), is presented in Fig.~\ref{Fig:Rb}(c). Our analysis indicates that in higher temperature regimes beyond $375\,$K, locking to a transmission maximum yields a higher sensitivity to temperature fluctuation. In fact, the utilization of transmission maxima can produce absorption changes exceeding $2\,\%/$K within this temperature range  where the transmission minima saturate to zero. However, when examining Fig.~\ref{Fig:Rb}(b), a region of reduced sensitivity to temperature change is apparent at approximately $400\,$K for both the transmission maxima and minima. 
Note that this analysis is based on only seven detuning values, and further investigation into alternative areas of the spectrum may yield more fruitful sensitivities within this range. 

In future work, we will conduct a more comprehensive analysis, encompassing a wider range of transitions, including the Rb D$_1$ line, as well as the D$_1$ and D$_2$ lines of other alkali metal species. This extended investigation aims to provide valuable insights into the most suitable atomic species for employing the proposed thermometry technique. Such findings will facilitate informed decision-making for future iterations of experimental testing.  Furthermore, future analysis will compare and contrast these findings with investigations of the equivalent changes in Doppler width per kelvin of temperature fluctuation. This analysis will serve as an important indicator for determining the most suitable atomic species for practical Doppler broadening thermometry applications. At present, our model does not encompass the effects of magnetic fields as well as collisional broadening and Dicke narrowing, which are expected to become increasingly significant at higher temperatures. 
It is therefore imperative to incorporate these additional factors into our investigations, to obtain a more comprehensive understanding of the behaviour and limitations of the optical transitions within the various alkali metal species -- guiding their selection for targeted applications. 


\maketitle

\section{Conclusion}

In conclusion, this study presents our initial findings in the development of practical Doppler broadening thermometry techniques. We have demonstrated the feasibility of deriving temperature readings from measurement of atomic absorption using the Rb D$_2$ line. Furthermore, we have identified the significant challenge to the fidelity of such measurements posed by the ASE originating from the laser diode during the spectroscopic analysis. 

Through the employment of narrow BW filters in the experimental setup and the characterization of the ASE using an optically dense vapour cell, as well as post processing techniques, we have effectively mitigated the adverse impact of the ASE. Our methodology has resulted in a more than 50-fold reduction of the ASE from $4.0\,$\% to $0.08\,$\%. Moreover, we have achieved an improvement in the theoretical agreement of the absorption spectra at room temperature, reducing the RMS error from $1.29\,$\% to  $0.33\,$\%. The dataset for this paper is available at [DOI].

For the purpose of open access, the author(s) has applied a Creative Commons Attribution (CC BY) licence to any Author Accepted Manuscript (AAM) version arising from this submission. We thank Aldo Mendieta of NPL for reviewing the original manuscript, as well as the thorough and valuable suggestions of the two anonymous referees.


\begin{thebibliography}{23}%
\makeatletter
\providecommand \@ifxundefined [1]{%
 \@ifx{#1\undefined}
}%
\providecommand \@ifnum [1]{%
 \ifnum #1\expandafter \@firstoftwo
 \else \expandafter \@secondoftwo
 \fi
}%
\providecommand \@ifx [1]{%
 \ifx #1\expandafter \@firstoftwo
 \else \expandafter \@secondoftwo
 \fi
}%
\providecommand \natexlab [1]{#1}%
\providecommand \enquote  [1]{``#1''}%
\providecommand \bibnamefont  [1]{#1}%
\providecommand \bibfnamefont [1]{#1}%
\providecommand \citenamefont [1]{#1}%
\providecommand \href@noop [0]{\@secondoftwo}%
\providecommand \href [0]{\begingroup \@sanitize@url \@href}%
\providecommand \@href[1]{\@@startlink{#1}\@@href}%
\providecommand \@@href[1]{\endgroup#1\@@endlink}%
\providecommand \@sanitize@url [0]{\catcode `\\12\catcode `\$12\catcode
  `\&12\catcode `\#12\catcode `\^12\catcode `\_12\catcode `\%12\relax}%
\providecommand \@@startlink[1]{}%
\providecommand \@@endlink[0]{}%
\providecommand \url  [0]{\begingroup\@sanitize@url \@url }%
\providecommand \@url [1]{\endgroup\@href {#1}{\urlprefix }}%
\providecommand \urlprefix  [0]{URL }%
\providecommand \Eprint [0]{\href }%
\providecommand \doibase [0]{http://dx.doi.org/}%
\providecommand \selectlanguage [0]{\@gobble}%
\providecommand \bibinfo  [0]{\@secondoftwo}%
\providecommand \bibfield  [0]{\@secondoftwo}%
\providecommand \translation [1]{[#1]}%
\providecommand \BibitemOpen [0]{}%
\providecommand \bibitemStop [0]{}%
\providecommand \bibitemNoStop [0]{.\EOS\space}%
\providecommand \EOS [0]{\spacefactor3000\relax}%
\providecommand \BibitemShut  [1]{\csname bibitem#1\endcsname}%
\let\auto@bib@innerbib\@empty
\bibitem [{\citenamefont {Fellmuth}\ \emph {et~al.}(2016)\citenamefont
  {Fellmuth}, \citenamefont {Fischer}, \citenamefont {Machin}, \citenamefont
  {Picard}, \citenamefont {Steur}, \citenamefont {Tamura}, \citenamefont
  {White},\ and\ \citenamefont {Yoon}}]{Fellmuth2016}%
  \BibitemOpen
  \bibfield  {author} {\bibinfo {author} {\bibfnamefont {B.}~\bibnamefont
  {Fellmuth}}, \bibinfo {author} {\bibfnamefont {J.}~\bibnamefont {Fischer}},
  \bibinfo {author} {\bibfnamefont {G.}~\bibnamefont {Machin}}, \bibinfo
  {author} {\bibfnamefont {S.}~\bibnamefont {Picard}}, \bibinfo {author}
  {\bibfnamefont {P.~P.~M.}\ \bibnamefont {Steur}}, \bibinfo {author}
  {\bibfnamefont {O.}~\bibnamefont {Tamura}}, \bibinfo {author} {\bibfnamefont
  {D.~R.}\ \bibnamefont {White}}, \ and\ \bibinfo {author} {\bibfnamefont
  {H.}~\bibnamefont {Yoon}},\ }\bibfield  {title} {\enquote {\bibinfo {title}
  {The kelvin redefinition and its \emph{mise en pratique}},}\ }\href
  {https://doi.org/10.1098/rsta.2015.0037} {\bibfield  {journal} {\bibinfo
  {journal} {Philos. Trans. Royal Soc. A}\ }\textbf {\bibinfo {volume} {374}},\
  \bibinfo {pages} {20150037} (\bibinfo {year} {2016})}\BibitemShut {NoStop}%
\bibitem [{\citenamefont {Preston-Thomas}(1990)}]{PrestonThomas1990}%
  \BibitemOpen
  \bibfield  {author} {\bibinfo {author} {\bibfnamefont {H.}~\bibnamefont
  {Preston-Thomas}},\ }\bibfield  {title} {\enquote {\bibinfo {title} {The
  international temperature scale of 1990 ({ITS}-90)},}\ }\href
  {https://doi.org/10.1088/0026-1394/27/1/002} {\bibfield  {journal} {\bibinfo
  {journal} {Metrologia}\ }\textbf {\bibinfo {volume} {27}},\ \bibinfo {pages}
  {3--10} (\bibinfo {year} {1990})}\BibitemShut {NoStop}%
\bibitem [{\citenamefont {Bramley}, \citenamefont {Cruickshank},\ and\
  \citenamefont {Aubrey}(2020)}]{Bramley2020}%
  \BibitemOpen
  \bibfield  {author} {\bibinfo {author} {\bibfnamefont {P.}~\bibnamefont
  {Bramley}}, \bibinfo {author} {\bibfnamefont {D.}~\bibnamefont
  {Cruickshank}}, \ and\ \bibinfo {author} {\bibfnamefont {J.}~\bibnamefont
  {Aubrey}},\ }\bibfield  {title} {\enquote {\bibinfo {title} {Developments
  towards an industrial {J}ohnson noise thermometer},}\ }\href
  {https://doi.org/10.1088/1361-6501/ab58a6} {\bibfield  {journal} {\bibinfo
  {journal} {Meas. Sci. Technol}\ }\textbf {\bibinfo {volume} {31}},\ \bibinfo
  {pages} {054003} (\bibinfo {year} {2020})}\BibitemShut {NoStop}%
\bibitem [{\citenamefont {Dedyulin}, \citenamefont {Ahmed},\ and\ \citenamefont
  {Machin}(2022)}]{Dedyulin2022}%
  \BibitemOpen
  \bibfield  {author} {\bibinfo {author} {\bibfnamefont {S.}~\bibnamefont
  {Dedyulin}}, \bibinfo {author} {\bibfnamefont {Z.}~\bibnamefont {Ahmed}}, \
  and\ \bibinfo {author} {\bibfnamefont {G.}~\bibnamefont {Machin}},\
  }\bibfield  {title} {\enquote {\bibinfo {title} {Emerging technologies in the
  field of thermometry},}\ }\href {\doibase 10.1088/1361-6501/ac75b1}
  {\bibfield  {journal} {\bibinfo  {journal} {Meas. Sci. Technol}\ }\textbf
  {\bibinfo {volume} {33}},\ \bibinfo {pages} {092001} (\bibinfo {year}
  {2022})}\BibitemShut {NoStop}%
\bibitem [{\citenamefont {{Daussy}}\ \emph {et~al.}(2005)\citenamefont
  {{Daussy}}, \citenamefont {{Briaudeau}}, \citenamefont {{Guinet}},
  \citenamefont {{Amy-Klein}}, \citenamefont {{Hermier}}, \citenamefont
  {{Bord{\'e}}},\ and\ \citenamefont {{Chardonnet}}}]{2005lasp.conf..104D}%
  \BibitemOpen
  \bibfield  {author} {\bibinfo {author} {\bibfnamefont {C.}~\bibnamefont
  {{Daussy}}}, \bibinfo {author} {\bibfnamefont {S.}~\bibnamefont
  {{Briaudeau}}}, \bibinfo {author} {\bibfnamefont {M.}~\bibnamefont
  {{Guinet}}}, \bibinfo {author} {\bibfnamefont {A.}~\bibnamefont
  {{Amy-Klein}}}, \bibinfo {author} {\bibfnamefont {Y.}~\bibnamefont
  {{Hermier}}}, \bibinfo {author} {\bibfnamefont {C.~J.}\ \bibnamefont
  {{Bord{\'e}}}}, \ and\ \bibinfo {author} {\bibfnamefont {C.}~\bibnamefont
  {{Chardonnet}}},\ }\bibfield  {title} {\enquote {\bibinfo {title}
  {Spectroscopic determination of the {B}oltzmann constant:. first results},}\
  }in\ \href {\doibase 10.1142/9789812701473_0010} {\emph {\bibinfo {booktitle}
  {Laser Spectroscopy}}}\ (\bibinfo {year} {2005})\ pp.\ \bibinfo {pages}
  {104--111}\BibitemShut {NoStop}%
\bibitem [{\citenamefont {Moretti}\ \emph {et~al.}(2013)\citenamefont
  {Moretti}, \citenamefont {Castrillo}, \citenamefont {Fasci}, \citenamefont
  {De~Vizia}, \citenamefont {Casa}, \citenamefont {Galzerano}, \citenamefont
  {Merlone}, \citenamefont {Laporta},\ and\ \citenamefont
  {Gianfrani}}]{PhysRevLett.111.060803}%
  \BibitemOpen
  \bibfield  {author} {\bibinfo {author} {\bibfnamefont {L.}~\bibnamefont
  {Moretti}}, \bibinfo {author} {\bibfnamefont {A.}~\bibnamefont {Castrillo}},
  \bibinfo {author} {\bibfnamefont {E.}~\bibnamefont {Fasci}}, \bibinfo
  {author} {\bibfnamefont {M.~D.}\ \bibnamefont {De~Vizia}}, \bibinfo {author}
  {\bibfnamefont {G.}~\bibnamefont {Casa}}, \bibinfo {author} {\bibfnamefont
  {G.}~\bibnamefont {Galzerano}}, \bibinfo {author} {\bibfnamefont
  {A.}~\bibnamefont {Merlone}}, \bibinfo {author} {\bibfnamefont
  {P.}~\bibnamefont {Laporta}}, \ and\ \bibinfo {author} {\bibfnamefont
  {L.}~\bibnamefont {Gianfrani}},\ }\bibfield  {title} {\enquote {\bibinfo
  {title} {Determination of the {B}oltzmann constant by means of precision
  measurements of {$\textrm{H}_{2}\,^{18}\textrm{O}$} line shapes at
  {$1.39\,\upmu\textrm{m}$}},}\ }\href
  {https://doi.org/10.1103/PhysRevLett.111.060803} {\bibfield  {journal}
  {\bibinfo  {journal} {Phys. Rev. Lett.}\ }\textbf {\bibinfo {volume} {111}},\
  \bibinfo {pages} {060803} (\bibinfo {year} {2013})}\BibitemShut {NoStop}%
\bibitem [{\citenamefont {Gianfrani}(2012)}]{Gianfrani2012}%
  \BibitemOpen
  \bibfield  {author} {\bibinfo {author} {\bibfnamefont {L.}~\bibnamefont
  {Gianfrani}},\ }\bibfield  {title} {\enquote {\bibinfo {title}
  {Highly-accurate line shape studies in the near-{IR} spectrum of
  {$\textrm{H}_2\,^{18}\textrm{O}$}: {I}mplications for the spectroscopic
  determination of the {B}oltzmann constant},}\ }\href
  {https://doi.org/10.1088/1742-6596/397/1/012029} {\bibfield  {journal}
  {\bibinfo  {journal} {J. Phys. Conf. Ser.}\ }\textbf {\bibinfo {volume}
  {397}},\ \bibinfo {pages} {012029} (\bibinfo {year} {2012})}\BibitemShut
  {NoStop}%
\bibitem [{\citenamefont {Gianfrani}(2016)}]{Gianfrani2016}%
  \BibitemOpen
  \bibfield  {author} {\bibinfo {author} {\bibfnamefont {L.}~\bibnamefont
  {Gianfrani}},\ }\bibfield  {title} {\enquote {\bibinfo {title} {Linking the
  thermodynamic temperature to an optical frequency: recent advances in
  {D}oppler broadening thermometry},}\ }\href {\doibase 10.1098/rsta.2015.0047}
  {\bibfield  {journal} {\bibinfo  {journal} {Philos. Trans. Royal Soc. A}\
  }\textbf {\bibinfo {volume} {374}},\ \bibinfo {pages} {20150047} (\bibinfo
  {year} {2016})}\BibitemShut {NoStop}%
\bibitem [{\citenamefont {Gotti}\ \emph {et~al.}(2018)\citenamefont {Gotti},
  \citenamefont {Moretti}, \citenamefont {Gatti}, \citenamefont {Castrillo},
  \citenamefont {Galzerano}, \citenamefont {Laporta}, \citenamefont
  {Gianfrani},\ and\ \citenamefont {Marangoni}}]{Gotti2018}%
  \BibitemOpen
  \bibfield  {author} {\bibinfo {author} {\bibfnamefont {R.}~\bibnamefont
  {Gotti}}, \bibinfo {author} {\bibfnamefont {L.}~\bibnamefont {Moretti}},
  \bibinfo {author} {\bibfnamefont {D.}~\bibnamefont {Gatti}}, \bibinfo
  {author} {\bibfnamefont {A.}~\bibnamefont {Castrillo}}, \bibinfo {author}
  {\bibfnamefont {G.}~\bibnamefont {Galzerano}}, \bibinfo {author}
  {\bibfnamefont {P.}~\bibnamefont {Laporta}}, \bibinfo {author} {\bibfnamefont
  {L.}~\bibnamefont {Gianfrani}}, \ and\ \bibinfo {author} {\bibfnamefont
  {M.}~\bibnamefont {Marangoni}},\ }\bibfield  {title} {\enquote {\bibinfo
  {title} {Cavity-ring-down {D}oppler-broadening primary thermometry},}\ }\href
  {\doibase 10.1103/physreva.97.012512} {\bibfield  {journal} {\bibinfo
  {journal} {Phys. Rev. A}\ }\textbf {\bibinfo {volume} {97}},\ \bibinfo
  {pages} {012512} (\bibinfo {year} {2018})}\BibitemShut {NoStop}%
\bibitem [{\citenamefont {Truong}\ \emph {et~al.}(2011)\citenamefont {Truong},
  \citenamefont {May}, \citenamefont {Stace},\ and\ \citenamefont
  {Luiten}}]{Truong2011}%
  \BibitemOpen
  \bibfield  {author} {\bibinfo {author} {\bibfnamefont {G.-W.}\ \bibnamefont
  {Truong}}, \bibinfo {author} {\bibfnamefont {E.~F.}\ \bibnamefont {May}},
  \bibinfo {author} {\bibfnamefont {T.~M.}\ \bibnamefont {Stace}}, \ and\
  \bibinfo {author} {\bibfnamefont {A.~N.}\ \bibnamefont {Luiten}},\ }\bibfield
   {title} {\enquote {\bibinfo {title} {Quantitative atomic spectroscopy for
  primary thermometry},}\ }\href {\doibase 10.1103/physreva.83.033805}
  {\bibfield  {journal} {\bibinfo  {journal} {Phys. Rev. A}\ }\textbf {\bibinfo
  {volume} {83}},\ \bibinfo {pages} {033805} (\bibinfo {year}
  {2011})}\BibitemShut {NoStop}%
\bibitem [{\citenamefont {Truong}\ \emph {et~al.}(2015)\citenamefont {Truong},
  \citenamefont {Stuart}, \citenamefont {Anstie}, \citenamefont {May},
  \citenamefont {Stace},\ and\ \citenamefont {Luiten}}]{Truong2015}%
  \BibitemOpen
  \bibfield  {author} {\bibinfo {author} {\bibfnamefont {G.-W.}\ \bibnamefont
  {Truong}}, \bibinfo {author} {\bibfnamefont {D.}~\bibnamefont {Stuart}},
  \bibinfo {author} {\bibfnamefont {J.~D.}\ \bibnamefont {Anstie}}, \bibinfo
  {author} {\bibfnamefont {E.~F.}\ \bibnamefont {May}}, \bibinfo {author}
  {\bibfnamefont {T.~M.}\ \bibnamefont {Stace}}, \ and\ \bibinfo {author}
  {\bibfnamefont {A.~N.}\ \bibnamefont {Luiten}},\ }\bibfield  {title}
  {\enquote {\bibinfo {title} {Atomic spectroscopy for primary thermometry},}\
  }\href {\doibase 10.1088/0026-1394/52/5/s324} {\bibfield  {journal} {\bibinfo
   {journal} {Metrologia}\ }\textbf {\bibinfo {volume} {52}},\ \bibinfo {pages}
  {S324--S342} (\bibinfo {year} {2015})}\BibitemShut {NoStop}%
\bibitem [{\citenamefont {Pan}\ \emph {et~al.}(2019)\citenamefont {Pan},
  \citenamefont {Liao}, \citenamefont {Wang}, \citenamefont {Yao},
  \citenamefont {Cai},\ and\ \citenamefont {Qu}}]{Pan:19}%
  \BibitemOpen
  \bibfield  {author} {\bibinfo {author} {\bibfnamefont {Y.}~\bibnamefont
  {Pan}}, \bibinfo {author} {\bibfnamefont {W.}~\bibnamefont {Liao}}, \bibinfo
  {author} {\bibfnamefont {H.}~\bibnamefont {Wang}}, \bibinfo {author}
  {\bibfnamefont {Y.}~\bibnamefont {Yao}}, \bibinfo {author} {\bibfnamefont
  {J.}~\bibnamefont {Cai}}, \ and\ \bibinfo {author} {\bibfnamefont
  {J.}~\bibnamefont {Qu}},\ }\bibfield  {title} {\enquote {\bibinfo {title}
  {Cesium atomic {D}oppler broadening thermometry for room temperature
  measurement},}\ }\href
  {https://opg.optica.org/col/abstract.cfm?URI=col-17-6-060201} {\bibfield
  {journal} {\bibinfo  {journal} {Chin. Opt. Lett.}\ }\textbf {\bibinfo
  {volume} {17}},\ \bibinfo {pages} {060201} (\bibinfo {year}
  {2019})}\BibitemShut {NoStop}%
\bibitem [{\citenamefont {Sherlock}\ and\ \citenamefont
  {Hughes}(2009)}]{Sherlock2009}%
  \BibitemOpen
  \bibfield  {author} {\bibinfo {author} {\bibfnamefont {B.~E.}\ \bibnamefont
  {Sherlock}}\ and\ \bibinfo {author} {\bibfnamefont {I.~G.}\ \bibnamefont
  {Hughes}},\ }\bibfield  {title} {\enquote {\bibinfo {title} {How weak is a
  weak probe in laser spectroscopy?}}\ }\href {\doibase 10.1119/1.3013197}
  {\bibfield  {journal} {\bibinfo  {journal} {Am.\ J.\ Phys.}\ }\textbf
  {\bibinfo {volume} {77}},\ \bibinfo {pages} {111--115} (\bibinfo {year}
  {2009})}\BibitemShut {NoStop}%
\bibitem [{\citenamefont {Zentile}\ \emph {et~al.}(2015)\citenamefont
  {Zentile}, \citenamefont {Keaveney}, \citenamefont {Weller}, \citenamefont
  {Whiting}, \citenamefont {Adams},\ and\ \citenamefont
  {Hughes}}]{Zentile2015}%
  \BibitemOpen
  \bibfield  {author} {\bibinfo {author} {\bibfnamefont {M.~A.}\ \bibnamefont
  {Zentile}}, \bibinfo {author} {\bibfnamefont {J.}~\bibnamefont {Keaveney}},
  \bibinfo {author} {\bibfnamefont {L.}~\bibnamefont {Weller}}, \bibinfo
  {author} {\bibfnamefont {D.~J.}\ \bibnamefont {Whiting}}, \bibinfo {author}
  {\bibfnamefont {C.~S.}\ \bibnamefont {Adams}}, \ and\ \bibinfo {author}
  {\bibfnamefont {I.~G.}\ \bibnamefont {Hughes}},\ }\bibfield  {title}
  {\enquote {\bibinfo {title} {{ElecSus}: A program to calculate the electric
  susceptibility of an atomic ensemble},}\ }\href
  {https://doi.org/10.1016/j.cpc.2014.11.023} {\bibfield  {journal} {\bibinfo
  {journal} {Comput. Phys. Commun.}\ }\textbf {\bibinfo {volume} {189}},\
  \bibinfo {pages} {162--174} (\bibinfo {year} {2015})}\BibitemShut {NoStop}%
\bibitem [{\citenamefont {Keaveney}, \citenamefont {Adams},\ and\ \citenamefont
  {Hughes}(2018)}]{Keaveney2018}%
  \BibitemOpen
  \bibfield  {author} {\bibinfo {author} {\bibfnamefont {J.}~\bibnamefont
  {Keaveney}}, \bibinfo {author} {\bibfnamefont {C.~S.}\ \bibnamefont {Adams}},
  \ and\ \bibinfo {author} {\bibfnamefont {I.~G.}\ \bibnamefont {Hughes}},\
  }\bibfield  {title} {\enquote {\bibinfo {title} {{ElecSus}: Extension to
  arbitrary geometry magneto-optics},}\ }\href
  {https://doi.org/10.1016/j.cpc.2017.12.001} {\bibfield  {journal} {\bibinfo
  {journal} {Comput. Phys. Commun.}\ }\textbf {\bibinfo {volume} {224}},\
  \bibinfo {pages} {311--324} (\bibinfo {year} {2018})}\BibitemShut {NoStop}%
\bibitem [{\citenamefont {Siddons}\ \emph {et~al.}(2008)\citenamefont
  {Siddons}, \citenamefont {Adams}, \citenamefont {Ge},\ and\ \citenamefont
  {Hughes}}]{Siddons2008}%
  \BibitemOpen
  \bibfield  {author} {\bibinfo {author} {\bibfnamefont {P.}~\bibnamefont
  {Siddons}}, \bibinfo {author} {\bibfnamefont {C.~S.}\ \bibnamefont {Adams}},
  \bibinfo {author} {\bibfnamefont {C.}~\bibnamefont {Ge}}, \ and\ \bibinfo
  {author} {\bibfnamefont {I.~G.}\ \bibnamefont {Hughes}},\ }\bibfield  {title}
  {\enquote {\bibinfo {title} {Absolute absorption on rubidium {D} lines:
  comparison between theory and experiment},}\ }\href
  {https://doi.org/10.1088/0953-4075/41/15/155004} {\bibfield  {journal}
  {\bibinfo  {journal} {J. Phys. B}\ }\textbf {\bibinfo {volume} {41}},\
  \bibinfo {pages} {155004} (\bibinfo {year} {2008})}\BibitemShut {NoStop}%
\bibitem [{\citenamefont {de~L.~Kronig}(1926)}]{deLKronig1926}%
  \BibitemOpen
  \bibfield  {author} {\bibinfo {author} {\bibfnamefont {R.}~\bibnamefont
  {de~L.~Kronig}},\ }\bibfield  {title} {\enquote {\bibinfo {title} {On the
  theory of dispersion of x-rays},}\ }\href {\doibase 10.1364/josa.12.000547}
  {\bibfield  {journal} {\bibinfo  {journal} {J. Opt. Soc. Am. A}\ }\textbf
  {\bibinfo {volume} {12}},\ \bibinfo {pages} {547} (\bibinfo {year}
  {1926})}\BibitemShut {NoStop}%
\bibitem [{\citenamefont {Steck}(2021{\natexlab{a}})}]{85RbSteck}%
  \BibitemOpen
  \bibfield  {author} {\bibinfo {author} {\bibfnamefont {D.~A.}\ \bibnamefont
  {Steck}},\ }\bibfield  {title} {\enquote {\bibinfo {title} {Rubidium 85 {D}
  line data},}\ }\href {http://steck.us/alkalidata} {\  (\bibinfo {year}
  {2021}{\natexlab{a}})}\BibitemShut {NoStop}%
\bibitem [{\citenamefont {Steck}(2021{\natexlab{b}})}]{87RbSteck}%
  \BibitemOpen
  \bibfield  {author} {\bibinfo {author} {\bibfnamefont {D.~A.}\ \bibnamefont
  {Steck}},\ }\bibfield  {title} {\enquote {\bibinfo {title} {Rubidium 87 {D}
  line data},}\ }\href {http://steck.us/alkalidata} {\  (\bibinfo {year}
  {2021}{\natexlab{b}})}\BibitemShut {NoStop}%
\bibitem [{\citenamefont {Schreier}(1992)}]{Schreier1992}%
  \BibitemOpen
  \bibfield  {author} {\bibinfo {author} {\bibfnamefont {F.}~\bibnamefont
  {Schreier}},\ }\bibfield  {title} {\enquote {\bibinfo {title} {The {V}oigt
  and complex error function: A comparison of computational methods},}\ }\href
  {\doibase 10.1016/0022-4073(92)90139-u} {\bibfield  {journal} {\bibinfo
  {journal} {J. Quant. Spectrosc. Radiat. Transf.}\ }\textbf {\bibinfo {volume}
  {48}},\ \bibinfo {pages} {743--762} (\bibinfo {year} {1992})}\BibitemShut
  {NoStop}%
\bibitem [{\citenamefont {Arnold}, \citenamefont {Wilson},\ and\ \citenamefont
  {Boshier}(1998)}]{Arnold1998}%
  \BibitemOpen
  \bibfield  {author} {\bibinfo {author} {\bibfnamefont {A.~S.}\ \bibnamefont
  {Arnold}}, \bibinfo {author} {\bibfnamefont {J.~S.}\ \bibnamefont {Wilson}},
  \ and\ \bibinfo {author} {\bibfnamefont {M.~G.}\ \bibnamefont {Boshier}},\
  }\bibfield  {title} {\enquote {\bibinfo {title} {A simple extended-cavity
  diode laser},}\ }\href {\doibase 10.1063/1.1148756} {\bibfield  {journal}
  {\bibinfo  {journal} {Rev.\ Sci.\ Instrum.}\ }\textbf {\bibinfo {volume}
  {69}},\ \bibinfo {pages} {1236--1239} (\bibinfo {year} {1998})}\BibitemShut
  {NoStop}%
\bibitem [{\citenamefont {Daffurn}, \citenamefont {Offer},\ and\ \citenamefont
  {Arnold}(2021)}]{Daffurn2021}%
  \BibitemOpen
  \bibfield  {author} {\bibinfo {author} {\bibfnamefont {A.}~\bibnamefont
  {Daffurn}}, \bibinfo {author} {\bibfnamefont {R.~F.}\ \bibnamefont {Offer}},
  \ and\ \bibinfo {author} {\bibfnamefont {A.~S.}\ \bibnamefont {Arnold}},\
  }\bibfield  {title} {\enquote {\bibinfo {title} {A simple, powerful diode
  laser system for atomic physics},}\ }\href {\doibase 10.1364/ao.426844}
  {\bibfield  {journal} {\bibinfo  {journal} {Appl. Opt.}\ }\textbf {\bibinfo
  {volume} {60}},\ \bibinfo {pages} {5832} (\bibinfo {year}
  {2021})}\BibitemShut {NoStop}%
\bibitem [{\citenamefont {Pizzey}\ \emph {et~al.}(2022)\citenamefont {Pizzey},
  \citenamefont {Briscoe}, \citenamefont {Logue}, \citenamefont
  {Ponciano-Ojeda}, \citenamefont {Wrathmall},\ and\ \citenamefont
  {Hughes}}]{Pizzey2022}%
  \BibitemOpen
  \bibfield  {author} {\bibinfo {author} {\bibfnamefont {D.}~\bibnamefont
  {Pizzey}}, \bibinfo {author} {\bibfnamefont {J.~D.}\ \bibnamefont {Briscoe}},
  \bibinfo {author} {\bibfnamefont {F.~D.}\ \bibnamefont {Logue}}, \bibinfo
  {author} {\bibfnamefont {F.~S.}\ \bibnamefont {Ponciano-Ojeda}}, \bibinfo
  {author} {\bibfnamefont {S.~A.}\ \bibnamefont {Wrathmall}}, \ and\ \bibinfo
  {author} {\bibfnamefont {I.~G.}\ \bibnamefont {Hughes}},\ }\bibfield  {title}
  {\enquote {\bibinfo {title} {Laser spectroscopy of hot atomic vapours: from
  'scope to theoretical fit},}\ }\href {\doibase 10.1088/1367-2630/ac9cfe}
  {\bibfield  {journal} {\bibinfo  {journal} {New J. Phys.}\ }\textbf {\bibinfo
  {volume} {24}},\ \bibinfo {pages} {125001} (\bibinfo {year}
  {2022})}\BibitemShut {NoStop}%
\end{thebibliography}

\providecommand{\noopsort}[1]{}\providecommand{\singleletter}[1]{#1}%

\end{document}